# rSHG: Re-scan Second Harmonic Generation Microscopy

**Stefan G. Stanciu,**[a, ‡,*] **Radu Hristu,**[a, ‡] **George A. Stanciu,**[a] **Denis E. Tranca,**[a] **Lucian Eftimie,**[b] **Adrian Dumitru,**[c,d] **Mariana Costache,**[c,d] **Harald A. Stenmark,**[e] **Harm Manders,**[f] **Amit Cherian,**[f] **Mariliis Tark-Dame,**[f] **and Erik M.M. Manders**[f]

[a]Center for Microscopy-Microanalysis and Information Processing, University Politehnica of Bucharest, 060042 Bucharest, Romania

[b]Emergency Military Hospital, Pathology Department, 010825 Bucharest, Romania

[c]Department of Pathology, Bucharest Emergency University Hospital, 050098 Bucharest, Romania

[d]Department of Pathology, Carol Davila University of Medicine and Pharmacy, Bucharest, Romania

[e]Department of Molecular Cell Biology, Institute for Cancer Research, Oslo University Hospital, Montebello, 0379 Oslo, Norway

[f]Confocal.nl, Science Park 106, 1098XG Amsterdam, The Netherlands

[‡]These authors contributed equally to this work

**\*Corresponding author:** Stefan G. Stanciu, E-mail: stefan.g.stanciu@upb.ro

**Abstract.** Second Harmonic Generation Microscopy (SHG) is generally acknowledged as a powerful tool for the label-free 3D visualization of tissues and advanced materials, with one of its most popular applications being collagen imaging. Although the great need, progress in super-resolved SHG imaging lags behind the developments reported over the past years in fluorescence-based optical nanoscopy. In this work, we quantitatively show on collagenous tissues that by combining SHG imaging with re-scan microscopy resolutions that surpass the diffraction limit with ~1.4x become available. Besides Re-scan Second Harmonic Generation Microscopy (rSHG), we demonstrate as well super-resolved Re-scan Two-Photon Excited Fluorescence Microscopy (rTPEF). These two techniques are implemented by modifying a Re-scan Confocal Microscope (RCM), retaining its initial function, resulting thus in a multimodal rSHG/rTPEF/RCM system. Given the simplicity and flexibility of re-scan microscopy, we consider that the reported results are likely to augment the number and nature of applications relying on super-resolved non-linear optical imaging.

**Keywords**: laser scanning microscopy, re-scan microscopy, second harmonic generation, two photon excited fluorescence, super-resolved optical imaging

# 1 Introduction

Nonlinear optical microscopies (NLO)[1-2] are currently regarded as valuable optical characterization tools, allowing high-resolution, label-free, 3D probing of important morpho-structural and chemical properties of bio-samples and advanced materials. Among these, Second Harmonic Generation Microscopy (SHG), exploits the interaction of two incident near-infrared (NIR) photons, usually generated by a femtosecond (fs) pulsed laser beam, with a non-centrosymmetric molecule, resulting in a single emitted photon with double the energy via a nonlinear process involving virtual states. Considering its high imaging potential based on intrinsic contrast, which is augmented by other notable advantages such as no bleaching, no blinking, or high signal to-noise ratio compared to fluorescence, SHG has gained important interest in the context of label-free characterization of fixed, ex-vivo and in-vivo tissues[3-4], and advanced materials[5-6]. With respect to the former, selective SHG contrast from collagen, the most abundant protein in the human body, holds valuable potential for tissue characterization and disease diagnostics[7-8], given that during the initiation and progression of a large variety of severe pathologies, including cancers[9-11], cardiovascular[12] or neurodegenerative[13] diseases, the collagen network of affected organs is modified in specific ways[14-15]. Tissue imaging applications based on SHG contrast from exogenous harmonophores are also noteworthy to mention[16-18]. In relationship to applications in materials science, SHG has been demonstrated as particularly useful for characterizing defects in semiconductors[19], carrier motion in organic transistors[20], grain boundaries for 2D materials[5], or for the overall better understanding of emerging materials[6, 21]. Third Harmonic Generation Microscopy (THG), achieved upon interaction of three-photons instead of two, originates from interfaces with sharp refractive index variations, and is also considered an important tool for label-free imaging[22].

Optical imaging at sub-diffraction resolutions is currently considered a high-interest topic, as progress in key scientific fields such as biology, medicine, materials science, or nanoelectronics depends on the spatial resolving power of the optical instruments used for characterization purposes. Fluorescence based super-resolved microscopy techniques offer typical resolutions in the range of 20-100nm[23-24]. Although they enable cutting edge bioimaging applications, even the most popular variants, based on pump & probe[24], or localization[23] strategies face important limitations due to the lack of chemical sensitivity and dependence on fluorescent probes, which can influence the behavior of live specimens which are imaged, while also holding cyto-toxicity implications. Furthermore, such issues may be accompanied by unpredictable anomalous processes related to the distribution of contrast agents in biological samples[25]. This motivates a strong need for super-resolution techniques that do not require fluorescent labels, but the current state-of-the art in far-field label-free imaging is yet to match the performances of fluorescence based optical nanoscopy techniques.

Among the first notable attempts to achieve super-resolved images based on harmonic generations have been reported by Masihzadeh[26] et al., and Liu et al.[27], who succeeded in reducing the size of the Point-Spread Function (PSF) by manipulating the polarization state of the incident light. While polarization-based strategies are very ingenious, they also present an important drawback in terms of high-sensitivity to effects such as circular dichroism or birefringence, which interfere with the polarization state of the excitation light. Later, Field et al.[28], introduced multiphoton spatial frequency-modulated imaging (MP-SPIFI), a technique providing super-resolved images in both SHG and Two-Photon Excited Fluorescence Microscopy (TPEF), which overcomes the above mentioned drawbacks related to specific polarization requirements. The authors experimentally showed a resolution improvement of 2x and demonstrated a theoretical limit of 4x. Although the great value of this work cannot be disputed, the intricate experimental setup raises some concerns on whether this method can be made available in easy-to-use systems that can be operated by a non-expert user. In an approach discussing a lesser resolution advantage under reduced technical complexity, Gregor et al.[29], reported a way to achieve super-resolved SHG and TPEF images, based on a system for image scanning microscopy (ISM) that can be easily obtained by lightly modifying a conventional setup for multiphoton microscopy. Under illumination with a 900 nm laser beam, the authors resolved for rat tail collagen fibrils distances of 550 nm with a contrast of 100%, which was not possible with conventional SHG imaging. ISM uses a mathematic algorithm on multiple (sometimes over $10^6$) small pinhole-images to reconstruct the image of the object with an improved resolution, which has implications with respect to the total acquisition time. More recent, Johnson et al. reported a ~2.3 x enhancement in SHG resolution[30] for collagen imaging, based on a photonic nanojet phenomena; this approach requires custom-synthesized microspheres to interact with the excitation light, which carries obvious limitations.

Herein, we demonstrate Re-scan Second Harmonic Generation Microscopy (rSHG), a new method for super-resolved SHG imaging, which overcomes many of the limitations discussed under the previous paragraph. We introduce as well Re-scan TPEF (rTPEF), with TPEF[31] being generally regarded as a "sister" NLO technique to SHG. Both rSHG and rTPEF rely on the Re-scan concept which stays at the core of Re-scan Confocal Microscopy (RCM)[32-34], a technique known to provide a ~1.4x resolution advantage over a conventional confocal microscope (with standard dyes), and 4x the sensitivity. In addition to a standard Confocal Laser Scanning Microscope (CLSM), where the beam is scanned on the sample by means of a system of mirrors, the RCM incorporates a second system of mirrors whose role is to scan the light emitted by the sample upon laser excitation on the surface of a CCD/s-CMOS image sensor. If the angular amplitude of the camera scanner (re-scanner) is double the angular amplitude of the sample scanner, a resolution advantage of √2 over a conventional CLSM is achieved. While other super-resolution techniques are known to provide higher resolution, such advantages come at the cost of high beam power, or very specific contrast agents[24]. Conversely, RCM's sensitivity allows using low, sample friendly, beam power levels (e.g. <20 kW/cm$^2$ instantaneous peak power density[35]), which is further augmented by the fact that the specifics of this technique accommodate any type of fluorophore. RCM is considered to be particularly useful for long-time imaging assays of live cells[35], where the possibility to use low beam power levels (even down to <1µW at the objective) can be of great benefit, but also for studies on fixed samples[36-38].

In this work we describe the architecture of a multimodal system that we have developed to accommodate the implementation of rSHG and rTPEF, next to the original RCM workmode, and provide a qualitative estimate and a quantitative assessment on the resolution advantage experimentally observed for rSHG on collagenous tissues, and for rTPEF on a fluorescent calibration sample. Our main focus of attention is placed on rSHG imaging, given the current high need for novel avenues for super-resolved label-free imaging.

## 2 Methods

### 2.1 Optical setup and data acquisition

rSHG and rTPEF were performed with a custom-developed multimodal prototype based on a conventional RCM setup of 1$^{st}$ generation, that was modified, **Fig. 1**. To enable rSHG and rTPEF imaging, the custom modified RCM unit was equipped with a free-space laser input port. This port was used to introduce in the original RCM excitation path[32] the fs-beam originating from an infrared (IR) Chameleon Vision II (Coherent) Ti-Sapphire laser with ~140 fs pulses, a repetition rate of 80 MHz, tuned at 860 nm. A dichroic mirror ZT775sp-2p (Chroma) (D1 in **Fig. 1**) was used to combine the continuous wave (CW) beam(s) used for RCM imaging (405nm, 488nm, 561nm, 633nm) with the fs-laser beam used for rSHG (860nm), to enable multimodal RCM/rSHG/rTPEF imaging in specific applications. By placing a mirror at the sample position, CW VIS light was "injected" on the fs IR arm. By adjusting the collinearity and overlap of the CW VIS and IR beams on the IR arm, outside the RCM case, allowed the alignment of the latter. The standard dichroic mirror and emission filter of a typical RCM unit have been kept in place, ZT405-488-561-640-NIR-rpc (Chroma) (D2 in **Fig. 1**) & ZET405-488-561-640m (Chroma) (F1 in **Fig. 1**), as they don't interfere with collecting the SHG/THG signals generated under 860nm fs IR beam excitation. Using the modified setup with other IR wavelengths for rSHG and rTPEF, may require removal of the emission and excitation filters used for standard RCM imaging. While the use of the pinhole is optional for rSHG and rTPEF, considering the intrinsic optical sectioning capabilities of these two techniques, removing it would interfere with the possibilities for multimodal RCM/rSHG/rTPEF imaging (not discussed here, in order not to divert focus). To avoid this, in the here reported setup, we used a pinhole that was kept fixed at 50 µm. Two additional short pass filters, ET750sp-2p8 (Chroma) (F2 &F3 in Fig. 1), were used to block the NIR excitation light. To extract SHG signals a band-pass filter ET430/24x (Chroma) (F4 in **Fig.** 1) was used, which was placed in a retractable 3D-printed filter holder, allowing switching between RCM, rSHG and rTPEF workmodes. For collecting the rTPEF images (on samples emitting no SHG signals) no band-pass filter was used.

The setup above described was coupled to a Ti2-E Nikon inverted microscope equipped with a 100x/1.42 NA objective. The rSHG and rTPEF images depicted in **Fig. 2**, **Fig. 3**, and **Fig. 4a)** were acquired with scientific-grade CMOS Hamamatsu Flash 4.0 V3 camera, 2048 X 2048 pixels, with a unit cell size of 6.5 μm (H) × 6.5 μm (V), operated in a non-cooled configuration. The use of a standard CMOS, Chameleon®3 5.0 MP CMOS camera with 2448 X 2048 pixels SONY IMX264 sensor, with a unit cell size of 3.45 μm (H) × 3.45 μm (V), was also evaluated, **Fig. 4b**).

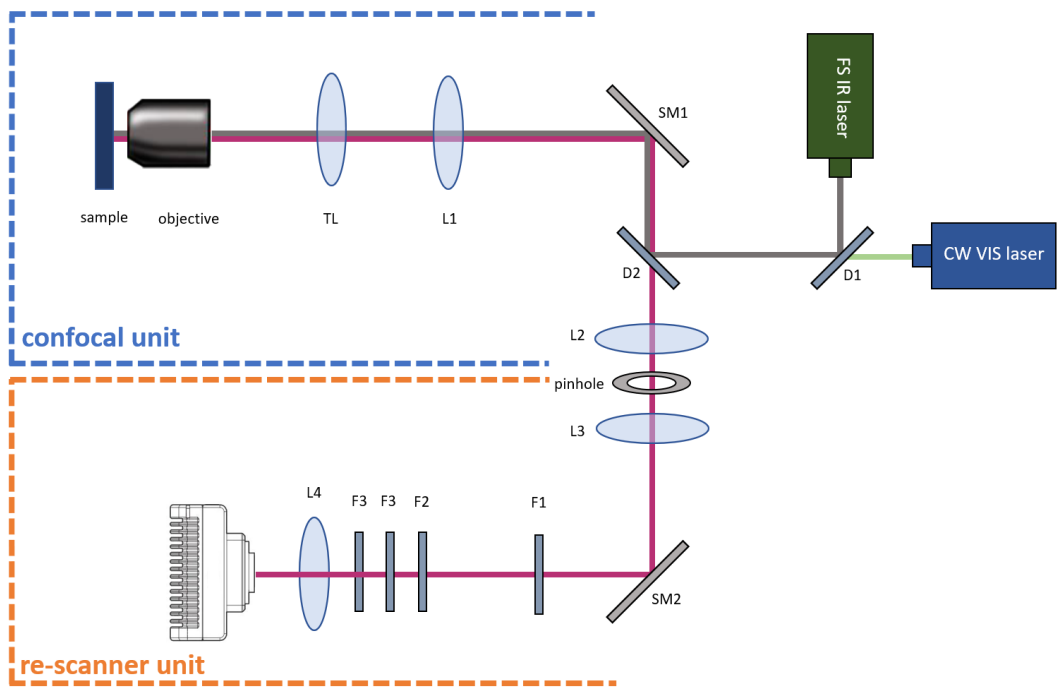

**Fig. 1**. Optical diagram of the prototype system for multimodal rSHG, rTPEF and RCM. The FS IR beam (grey) is inserted in the RCM default CW vis illumination light path with a beam combiner (D1). A second dichroic mirror (D2) inserts the laser beam(s) in the re-scan light path and separates the excitation from emission. It forwards the laser beam(s) to a first pair of scanning mirrors (SM1) which accomplish point-by-point sample excitation. The light emitted by the sample (violet) is de-scanned by the same SM1, which directs it via a pinhole to the second pair of scanning mirrors (SM2). The light rescanned by SM2 is passed through a quad band emission filter (F1) tuned to the typical emissions range for RCM fluorescence imaging, two NIR blockers (F2 &F3) suppress unwanted contributions from the fundamental FS IR beam, a bandpass filter (F4) is used to extract the rSHG signals of interest. For rTPEF this filter can be replaced with a long pass filter. For combined rSHG and rTPEF imaging, this filter position can be left empty. The emitted light finally reaches the camera after being collected by a lens (L4). The resolution advantage of rSHG and rTPEF is available when the angular amplitude of SM2 is the double of SM1's. [TL-Tube Lens; L-Lens; SM-Scanning Mirrors].

## 2.2 Image acquisition and registration

In a typical RCM system, the CW laser is switched off during the return movement of the scanner to avoid ghosting effects (blur) that typically arise due to the mismatch in the angular speed of the scanner and re-scanner, during their forward and backward movement. This leads to a phase difference of the scanner relative to the re-scanners. This problem is normally solved by hardware triggering, which synchronizes the on/off status of the laser with the forward/backward movement of the scanner. Implementing a similar strategy with the NIR fs-laser beam was not feasible as such high-frequency on/off modulation interferes with the pulse locking. Therefore, we used bi-directional scanning with a phase shift correction to minimize ghost image effects. Alternatively, a hardware-triggered Acusto-Optical Modulator positioned on the path of the NIR fs-laser beam can potentially be used to meet the same objective.

As thoroughly described in the work of De Luca et al.[32] improved resolution with the Re-scan amplitude of the camera scanner (re-scanner) is double the angular amplitude of the sample scanner (concept is achieved when the angular Sweep factor [SF] = 2, in Confocal.nl RCM terminology); the resolution available when the amplitudes of the two scanners match (SF = 1) is similar to the resolution of a conventional, diffraction limited, confocal laser scanning microscope. In **Fig. 2, Fig. 3** and **Fig. 4** we present rSHG images collected at SF = 1 and SF = 2, which we denote as SHG and rSHG images, respectively. A similar naming convention is used for the rTPEF/TPEF images. The rSHG images displayed in **Fig. 2**, **Fig. 3** and **Fig. 4a**), together with the rTPEF images presented in **Fig. 3**, were collected with the Hamamatsu ORCA Flash4.0 v3 camera, with 43nm pixel size, whereas the SHG and TPEF images in these figures were collected with 86nm pixel size (a consequence of the SF setting). The rSHG and SHG images collected with the Chameleon®3 camera, **Fig. 4b**) have been collected with 23nm and 46 nm pixel size, respectively.

Considering the difference in pixel size, to observe the same sample regions in both rSHG and SHG configurations, we used digital resolutions of 1024x1024 pixels and 512x512 pixels, respectively. The 1024x1024 pixels rSHG images were collected with a scanning speed of 200 Hz, and the same pixel dwell time was used for the SHG images. For qualitative assessment of the resolution advantage, we registered the SHG images to the rSHG ones by 2x digital upsampling (no interpolation). For the quantitative assessment part, all calculations and profile lines were measured on the original (non-registered and non-processed) images. The same methodology was used for the acquisition and analysis of the rTPEF/TPEF images.

## 2.3 Sample preparation

*Preparation of the rat-tail tendon samples:* The imaged rat-tail tendon tissues were fixed with 10% buffered formalin for 24 hours and were processed by conventional histopathological methods using paraffin embedding, 3 μm thick sectioning. These samples have not been stained.

*Preparation of the Hematoxylin–Eosin (HE) stained breast tissue samples:* The tissue fragments discussed in this work were previously collected for routine histological exam from a patient diagnosed with moderately differentiated (G2) invasive breast cancer of no special type (NST) with extensive foci of high grade ductal carcinoma in situ (DCIS), using the current criteria from WHO Classification of Tumors of the Breast, 5th edition[39]. The tumor grade was established using the Elston-Ellis (Nottingham score) grading system. The same samples previously used for clinical histological exam were imaged with rSHG/SHG (**Fig. 4** of the manuscript). Specifically, rSHG was used to image normal breast tissue from healthy regions close to the resection margins of malignant/premalignant lesions. The imaged specimen samples were fixed with 10% buffered formalin for 24 hours and were processed by conventional histopathological methods using paraffin embedding, 3 $\mu$m thick sectioning and HE staining.

## 3  Results & discussions

For demonstrating the resolution advantage of rSHG we have first used a rat-tail tissue fragment (fixed, unlabeled), a highly collagenous tissue that is emerging as a model for benchmarking SHG imaging methods. The importance of SHG imaging for characterizing the collagen architecture in tissues has been widely discussed to date[3, 40-41]. In brief, given that collagen is the main component in the extracellular matrix of mammalian tissues, SHG imaging of this protein with non-centrosymmetric structure provides important information on how the structure of tissues is modified along the progression of various pathologies including cancers[3, 7, 14]. In **Fig. 2** we provide qualitative evidence on the resolution advantage of rSHG over conventional, diffraction limited SHG imaging. The yellow arrows point to sample areas where the collagen fibers, and the overall aspect of the collagen network is significantly better resolved in the rSHG images compared to the SHG ones.

In **Fig. 3** we provide quantitative evidence on the resolution enhancement provided by the two considered NLO techniques, rSHG and rTPEF. All calculations discussed in this figure were performed on non-processed, non-averaged, rSHG and rTPEF images. In **Fig. 3a)** we present a rat tail tissue collagenous sample area where profile lines were drawn over well separated collagen fibrils. Besides the better resolved single collagen fibrils in the middle of the imaged region, the resolution advantage of rSHG is also very well visible for the fibril bundles in the bottom left and top right part of the image. In **Fig. 3b)** we present a close-up view of a single collagen fibril, highlighting a region where the SHG and rSHG profile lines were drawn to quantitatively assess the resolution enhancement. The average the Full-Width Half Maximum (FWHM) ratio for 10 profile lines randomly drawn over the highlighted area was 1.38±0.23, with the average FWHM calculated for the rSHG and SHG images being 160±16 and 222±29, respectively (results presented as Mean±SD). Similar values were identified for >70 considered profile lines drawn over single collagen fibrils in other acquired SHG & rSHG image pairs.

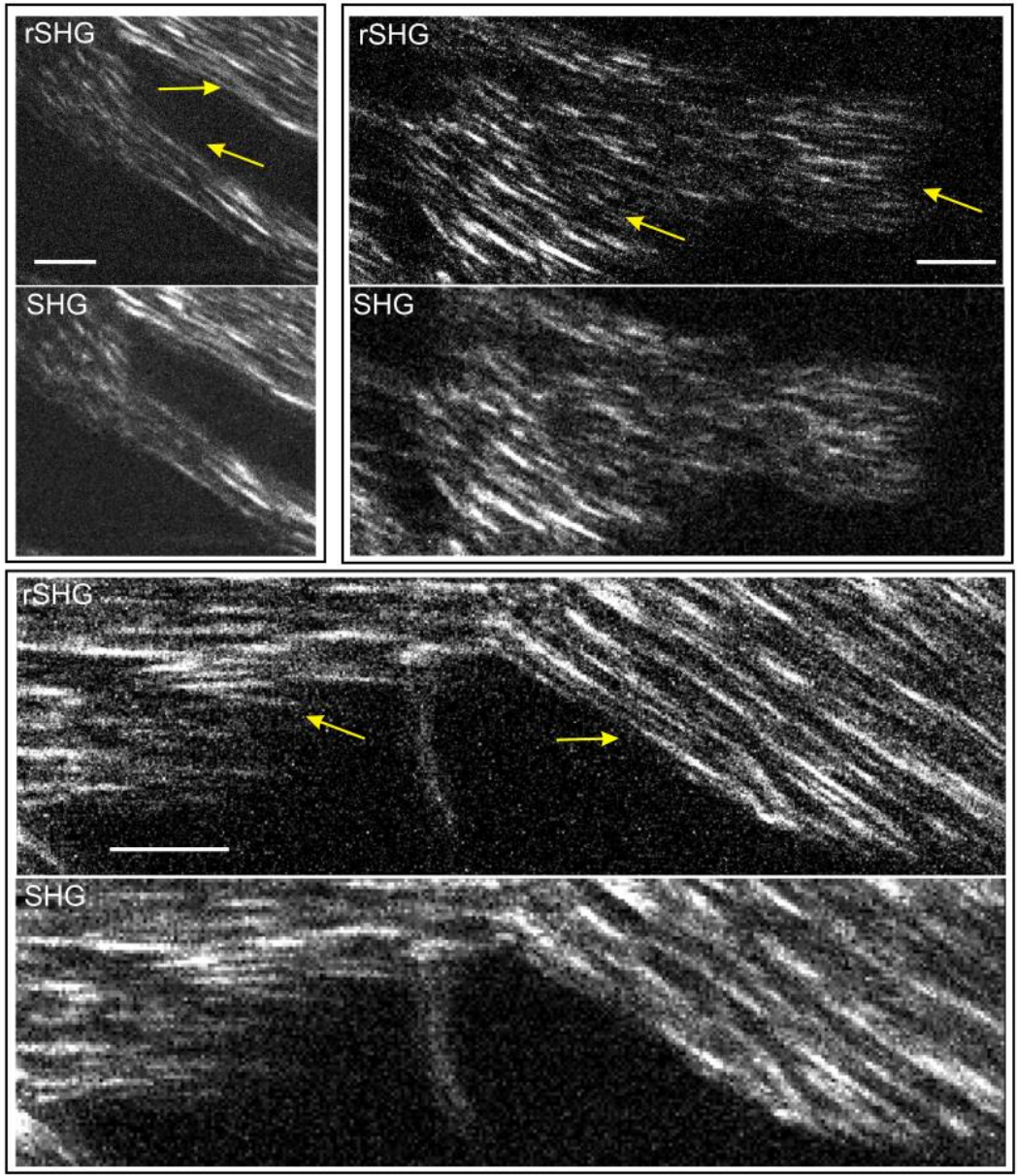

**Fig. 2**. Resolution advantage of rSHG vs. SHG on rat-tail collagenous tissues at different levels of detail. Yellow arrows point to sample areas where rSHG's resolving power is well visible. Scale bars: 3μm.

Noteworthy, the experimental FWHM values corresponding to the SHG image align to the theoretical value, which for 860nm illumination wavelength and an objective of 1.42 NA is 210 nm, according to Eq. 1[42]:

$$\text{FWHM} = \frac{0.5\lambda}{\sqrt{2} \cdot \eta \cdot sin\theta} \qquad \text{Eq. 1}$$

With the same setup we also explored rTPEF (**Fig. 3c**), which was performed on a fluorescent Argo SIM (Argolight, France) calibration slide. For >50 profile lines pairs randomly considered in the displayed TPEF and rTPEF images, the average FWHM ratio was 1.51±0.05, with the average FWHM calculated for the rTPEF and TPEF images being 241±5 and 364±10, respectively. Examples of TPEF & rTPEF profile lines are provided in **Fig. 3d**.

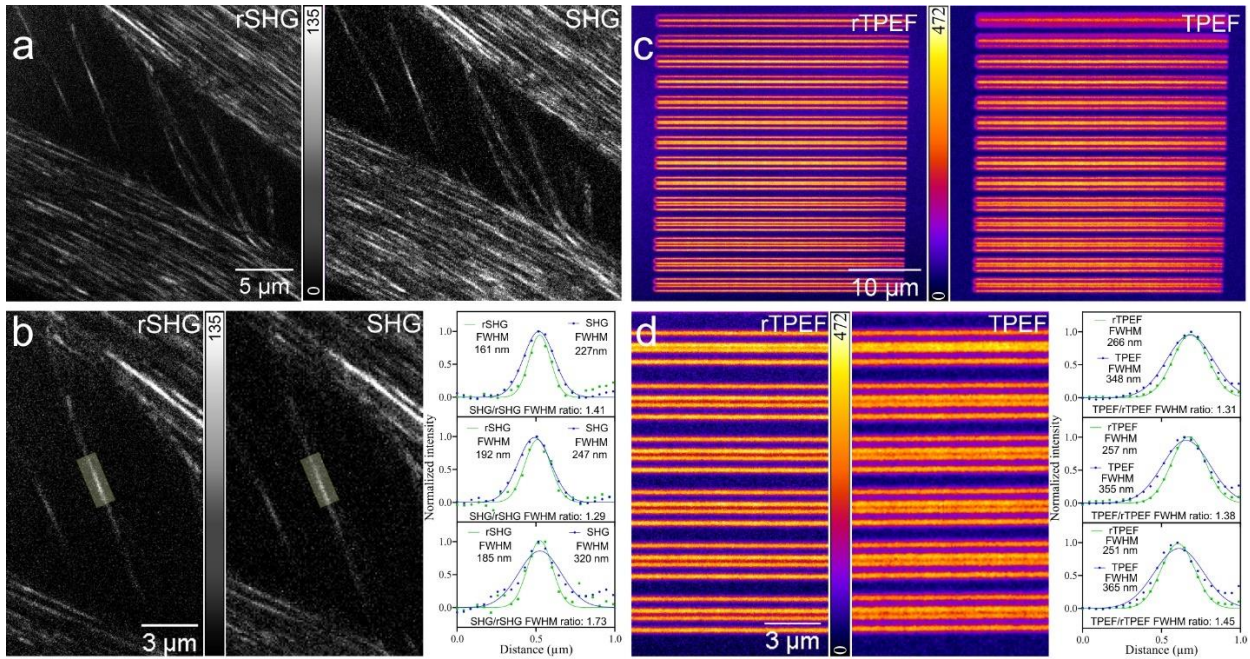

**Fig. 3**. Resolution improvement in rSHG and rTPEF. a) rSHG and SHG images collected on rat tail tendon collagen fibrils. b) Examples of profile line pairs transversally drawn across the collagen fibril in the highlighted area (green: rSHG, blue: SHG). c) rTPEF & TPEF images collected on an Argo SIM calibration sample. d) Examples of profile line pairs transversally drawn across the visible structures (green-rTPEF, blue-TPEF).

The resolution advantage of rSHG can also be observed in the case of a practical application, referring to the assessment of human breast tissue samples. In **Fig. 4** we present Re-scan and diffraction limited SHG images collected on a H&E-stained normal breast tissue fragment, prepared according to typical procedures for clinical histopathological exam. As discussed in past work, H&E staining does not significantly influence SHG signals[43]. The rSHG images displayed in **Fig. 4** represent cropped regions of rSHG/SHG mosaics that were assembled from 2x2 image tiles with the MosaicJ[44] plugin of ImageJ[45]. For the rSHG images collected with the standard CMOS Chameleon®3 camera (**Fig. 2b**), hot pixels were removed with the standard despeckle denoising option of Fiji[46], whereas no image processing has been performed for the rSHG images collected with the scientific grade CMOS Hamamatsu ORCA-Flash4.0 v3 camera. In **Fig. 4**, it can be noticed that rSHG allows assessing the conformation and disposition of the collagen in the connective stroma neighboring a healthy mammary terminal ductal-lobular unit (according to registered H&E images, not shown here), with considerable better clarity than SHG. Although discussing in detail the histopathological interpretation of this images lies outside the scope of this paper, we note that the superior resolution available with rSHG allows easier visual segmentation of various features of interest, such as individual collagen fibrils, or fibril formations. Furthermore, the greater level of detail can obviously favor the more accurate extraction of quantitative information reflecting the morphology and other properties of the collagen fibrils[47-50], which play an important role in computer-aided diagnostics scenarios[51]. Furthermore, it is important to highlight that for methods aimed at differentiating pathological vs. healthy tissues based on pixel-by-pixel fitting of polarization-resolved SHG images of collagen fibrils[52-53],[41] a higher optical resolution (hence lower pixel size) enables access to additional information which can augment the diagnostics procedure.

In terms of relevant applications, given that an RCM unit adapted for rSHG can be easily coupled to optical microscopes already available in clinical settings, we argue that histopathology labs can benefit from this reported methodology, as it can enable modern workflows based on

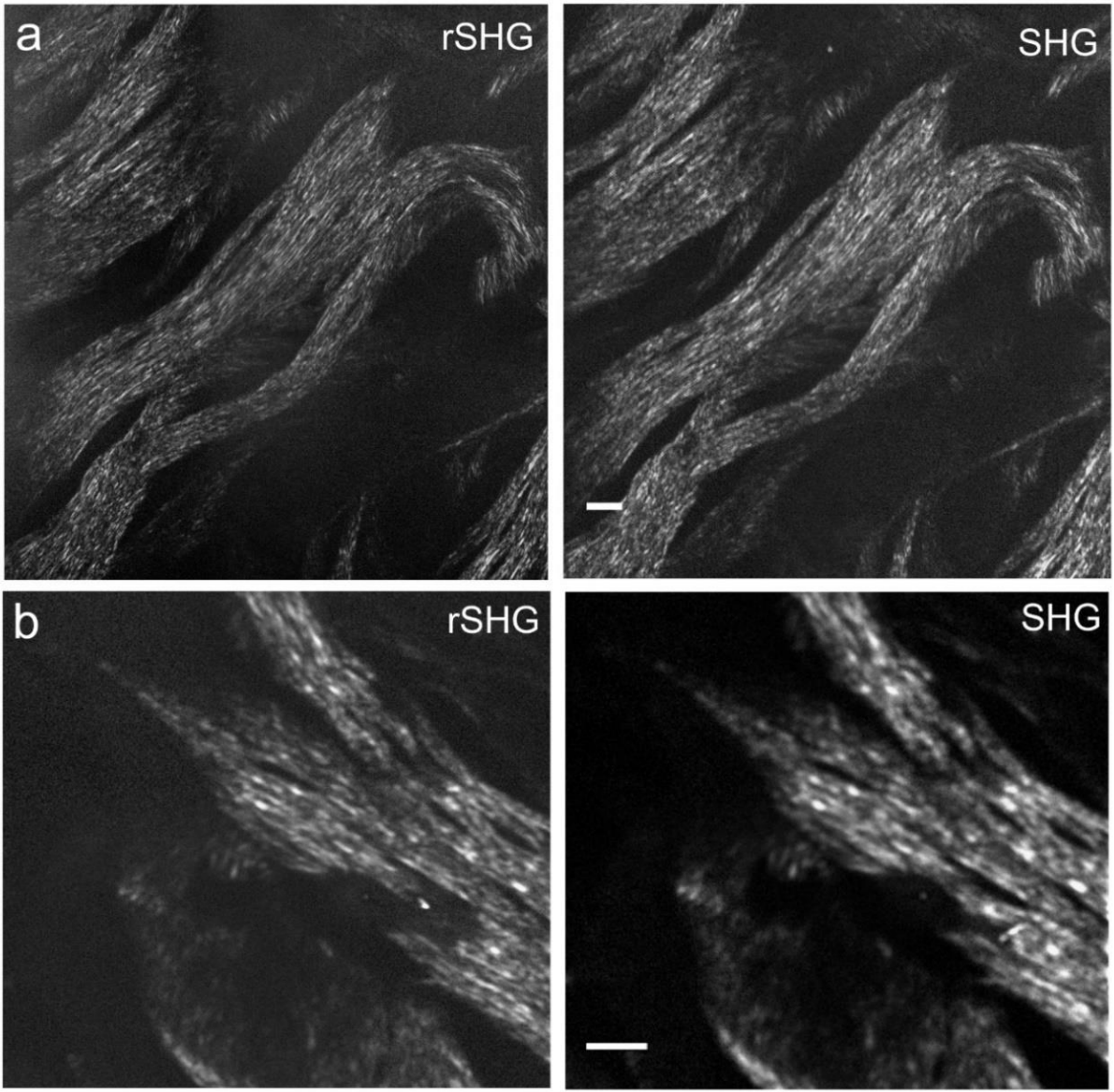

**Fig. 4**. SHG and rSHG images of a tissue fragment from a negative resection margin adjacent to a breast carcinoma acquired with a) a scientific grade Hamamatsu ORCA-Flash4.0 v3 s-CMOS, and b) a standard CMOS Chameleon®*3* camera. Scale bar: 5μm.

NLO techniques to complement traditional assays. Such correlative approaches can be implemented on samples prepared using traditional histopathology protocols, given that usual dyes (e.g., H&E), do not significantly affect the SHG signals. On the other hand, such stains have been shown to augment THG tissue imaging[54], which can potentially be implemented with the same

Re-scan architecture, depending on the availability of appropriate illumination sources. Furthermore, considering that super-resolved imaging of collagen, and various endogenous chromophores, can also be performed on unstained tissues with rSHG and rTPEF, respectively, we argue that these techniques can be of important help to identify subtle modifications that precede, or develop, during cancers on freshly excised tissues, next to the patient. Using these either separately or in-tandem could significantly promote improved diagnostics and faster decision making. Such potential approaches can be assisted by NLO-oriented artificial intelligence methods[10, 55]. The utility of multimodal rSHG/rTPEF/RCM systems, can be eventually extended by incorporating additional workmodes sharing common principles such as Re-scan THG, or Re-scan Coherent anti-Stokes Raman scattering microscopy, which would provide valuable complementary information on the structural and biochemical configuration of the imaged tissues, and of bio-samples in general[22, 56-58].

## 3    Conclusions

We report a multimodal prototype system that exploits the concepts of the Re-Scan Confocal Microscope to enable super-resolved rSHG and rTPEF imaging, next to the original RCM workmode. With this imaging system we show that the advantages of RCM over conventional CLSM, such as increased resolution and sensitivity are also available with the new Re-scan workmodes, demonstrating the versatility of this technology, and its potential to enable a new wide range of applications focused on super-resolved label-free and label-based NLO imaging. The resolution improvement of rSHG, that we experimentally assessed as 1.38±0.23 over diffraction limited SHG, was shown to result in images with considerable better clarity on the collagenous tissues that we investigated, providing access to subtle details that can significantly augment tissue diagnostics. The same was observed for rTPEF images collected on a fluorescent calibration sample, for which we experimentally demonstrated a 1.51±0.05 resolution improvement over diffraction limited TPEF. Considering the reduced complexity of Re-scan setups, we argue that this work will contribute to the widespread availability of super-resolved SHG and TPEF imaging, and of NLO imaging, in general.

*Disclosures*

HM, MTD, AC, and EM disclose financial interest in Confocal.nl. EM is co-inventor on several patents (NL2026002B1, NL2023384B1, NL2020516B1, NL2018134B1) related to the re-scan confocal microscopy, which lies at the foundation of the here reported work. The remaining authors declare no competing interests.

*Acknowledgments*

This work was supported by the Horizon 2020 Attract: HARMOPLUS Project (GA 777222, project Id: 1052), and by the Romanian Executive Agency for Higher Education, Research, Development and Innovation Funding (UEFISCDI) via grants: RO-NO-2019-0601 and PN-III-P2-2.1-PED-2019-1666. SGS expresses sincere thanks to Dr. Arnaud Royon at Argolight for providing the ARGO SIM calibration slide for testing.

*Author contributions*

The manuscript was written through contributions of all authors. All authors have given approval to the final version of the manuscript.

*References*

(1) Mazumder, N.; Balla, N. K.; Zhuo, G.-Y.; Kumar, R.; Kao, F.-J.; Kistenev, Y. V.; Brasselet, S.; Nikolaev, V. V.; Krivova, N. A. Label-free Nonlinear Multimodal Optical Microscopy-Basics, Development and Applications. *Frontiers in Physics* **2019,** *7*, 170.

(2) Zipfel, W. R.; Williams, R. M.; Webb, W. W. Nonlinear magic: multiphoton microscopy in the biosciences. *Nature biotechnology* **2003,** *21* (11), 1369.

(3) Campagnola, P. J.; Dong, C. Y. Second harmonic generation microscopy: principles and applications to disease diagnosis. *Laser & Photonics Reviews* **2011,** *5* (1), 13-26.

(4) Dilipkumar, A.; Al-Shemmary, A.; Kreiß, L.; Cvecek, K.; Carlé, B.; Knieling, F.; Gonzales Menezes, J.; Thoma, O. M.; Schmidt, M.; Neurath, M. F. Label-Free Multiphoton Endomicroscopy for Minimally Invasive In Vivo Imaging. *Advanced Science* **2019**, 1801735.

(5) Karvonen, L.; Säynätjoki, A.; Huttunen, M. J.; Autere, A.; Amirsolaimani, B.; Li, S.; Norwood, R. A.; Peyghambarian, N.; Lipsanen, H.; Eda, G. Rapid visualization of grain boundaries in monolayer MoS 2 by multiphoton microscopy. *Nature communications* **2017,** *8* (1), 1-8.

(6) Gleeson, M.; O'Dwyer, K.; Guerin, S.; Rice, D.; Thompson, D.; Tofail, S. A.; Silien, C.; Liu, N. Quantitative Polarization-Resolved Second-Harmonic-Generation Microscopy of Glycine Microneedles. *Advanced Materials* **2020,** *32* (46), 2002873.

(7) Ouellette, J. N.; Drifka, C. R.; Pointer, K. B.; Liu, Y.; Lieberthal, T. J.; Kao, W. J.; Kuo, J. S.; Loeffler, A. G.; Eliceiri, K. W. Navigating the collagen jungle: the biomedical potential of fiber organization in cancer. *Bioengineering* **2021,** *8* (2), 17.

(8) Campagnola, P., Second harmonic generation imaging microscopy: applications to diseases diagnostics. ACS Publications: 2011.
17